\pdfoutput=1
\documentclass[HYPER]{JHEP}
\usepackage{graphicx,amssymb,bm,latexsym,amsmath,mathtools}
\usepackage{epstopdf}
\usepackage{caption}
\usepackage{subcaption}
\usepackage[section]{placeins}
\usepackage{float}
\usepackage{capt-of}
\usepackage{tabu}
\usepackage[toc,page]{appendix}
\usepackage[toc,page]{appendix}
\usepackage[utf8]{inputenc}
\DeclarePairedDelimiter\abs{\lvert}{\rvert}%
\title{On finite-size spiky strings in $AdS_3 \times S^3\times T^4$ with mixed fluxes}
\author{Sorna Prava Barik, Rashmi R. Nayak and Kamal L. Panigrahi \\
	Department of Physics, Indian Institute of Technology Kharagpur,
	Kharagpur-721 302, India\\
	Email: \email{sornaprava15@gmail.com,rashmi.string@gmail.com, panigrahi@phy.iitkgp.ac.in}}

\abstract{We discuss finite-size corrections to the spiky strings in $AdS$ space which is dual to the long $\mathcal{N}=4$ SYM operators of the form Tr($\Delta_+ ^{J_1}\phi_1\Delta_+ ^{J_2}\phi_2...\Delta_+ ^{J_n}\phi_n$). We express the finite-size dispersion relation in terms of Lambert $\mathbf{W}$-function. We further establish the finite-size scaling relation between energy and angular momentum of the spiky string in  presence of mixed fluxes in terms of $\mathbf{W}$-function. We comment on the solution in pure NS-NS background as well.}
\begin{document}
	\section{Introduction}
AdS/CFT correspondence relates type IIB superstring theory on $AdS_5\times S^5$ background and the $\mathcal{N}=4$ supersymmetric Yang-Mills theory in flat space $\mathbb{R}^{1,3}$ with gauge group $SU(N)$ \cite{Maldacena:1997re, Witten:1998qj, Gubser:1998bc}. This holographic conjecture asserts one-to-one correspondence between various aspects of two models (e.g. global symmetries, spectra, correlation functions, Wilson loops, scattering amplitudes etc.). However, the precise matching of spectra on both 
sides of duality is really challenging and it is only recently, the problem has become tractable in 
infinite or large charge limit. The most remarkable development in this direction was made by Gubser, Klebanov and
Polyakov (GKP) in 2002 [4], where they studied the closed and folded string spinning in $AdS_3\subset AdS_5$. The folded spinning string dispersion relation for large angular momentum was found out to be,
\begin{eqnarray}
E-J=\frac{\sqrt{\lambda}}{\pi} \ln\frac{J}{\sqrt{\lambda}},\qquad \sqrt{\lambda}\gg 1.
\end{eqnarray}
The important observation they made was that the difference between the energy and the
angular momentum of the long string $(E-J)$, scales as the logarithm of the angular momentum $(\ln J)$, which was very similar to the anomalous dimension of twist two Wilsonian operators in perturbative QCD. It has been shown that this
string configuration is dual to the twist two operators of $\mathcal{N}=4$ SYM theory which has the
form Tr$(\phi\Delta_+^J\phi)$\footnote{Here the field $\phi$ is one of the adjoint complex scalars in 
super Yang-Mills theory and $\Delta_+$ denotes the covariant derivative in light cone coordinates.}. This observation made the 
single trace operators in large $N$ gauge theory relevant in  the AdS/CFT correspondence. Thereafter
the twist two operator has been generalised to higher twist operator with arbitrary number of fields 
of the form
\begin{eqnarray}
\text{Tr}(\Delta_+^{J_1}\phi_1\Delta_+^{J_2}\phi_2...\Delta_+^{J_n}\phi_n).
\end{eqnarray}
In large spin limit, the anomalous dimension of this operator is completely dominated by the
contribution from derivatives. The string solution dual to such a operator has been discussed in 
\cite{Kruczenski:2004wg} for  $J_1=J_2=...= \frac{J}{n}$. The dispersion relation of dual string 
which has $n$ spikes approaching the boundary is given by
\begin{eqnarray}
E-J=\frac{n\sqrt{\lambda}}{2\pi} \ln\left(\frac{4\pi J}{n\sqrt{\lambda}}\right),~~ J\rightarrow\infty.
\end{eqnarray}
A lot of activities have followed this work, e.g.
\cite{Kruczenski:2006pk,Kruczenski:2008bs,Jevicki:2008mm,Ishizeki:2008tx,Jevicki:2009uz,Kruczenski:2010xs,Banerjee:2015nha} and further, the spin chain connection have been studied in detail, e.g. 
\cite{Dorey:2008vp,Dorey:2010iy,Dorey:2010id,Dorey:2010zz}.\par
\quad Apart from most studied $AdS_5/CFT_4$ duality, the other holographic set up which are amenable 
to integrability is $AdS_3/CFT_2$. In this case, the type IIB string theory on $AdS_3\times S^3\times 
T^4$ is shown to be dual to the $\mathcal{N}=(4,4)$ superconformal field theory in two dimensions. Several semiclassical string solutions have been investigated in  $AdS_3\times S^3\times T^4$ geometry which arises from the near horizon limit of the intersecting D1-D5 branes in supergravity. The superstring theory on $AdS_3\times S^3$ geometry 
supported by both NS-NS and RR fluxes has  been proven to be integrable and the S-matrix has been 
constructed for the same 
\cite{Hoare:2013pma,Hoare:2013ida,OhlssonSax:2018hgc,Lloyd:2014bsa,Borsato:2015mma,Baggio:2018gct}\footnote[2]{see 
\cite{Sfondrini:2014via} for a review and references therein to get the detailed approach to this problem.}.  Recently, mirror
thermodynamic Bethe ansatz (TBA) has been proposed in pure NS-NS $AdS_3\times S^3\times S^3\times S^1 $ and $AdS_3\times S^3\times T^4 $ backgrounds \cite{Dei:2018jyj,Dei:2018mfl}. And also, the TBA has been contructed for low-energy string excitations as well as massless modes in RR background  \cite{Bombardelli:2018jkj,Fontanella:2019ury}. In \cite{Pittelli:2014ria,Pittelli:2017spf}, the complete Yangian symmetry underlying the integrability of type IIB superstrings on $AdS_3 \times S^3 \times T^4$ and $AdS_3\times S^3\times S^3\times S^1$ supported by mixed R-R and NS-NS fluxes have been derived. The giant magnon solution in mixed fluxes background was studied in \cite{Hoare:2013lja} followed by many other semiclassical string solutions 
\cite{Lee:2008sk,David:2014qta,Banerjee:2014gga,Banerjee:2015qeq}.\par
In the context of AdS/CFT correspondence, it has long been known that the gauge theory as well as 
string theory are integrable in the limit of very large or infinite global charges. 
The dilatation operators of the $\mathcal{N}=4$ SYM theory acting on the gauge invariant single trace
operators can be mapped to the Hamiltonian of an integrable spin chain model which can be
diagonalized by Bethe ansatz \cite{Minahan:2002ve,Beisert:2003jj,Beisert:2003yb,Beisert:2004hm}. 
The asymptotic Bethe ansatz predicts the correct form of the anomalous dimension up to the order of
$\lambda^L$ only when the length of the spin chain ($L$) is infinite or larger than the loop order. 
However, when the range of spin chain interaction exceeds the length of spin chain, the virtual 
particles start circulating around the spin chain giving rise to wrapping effect. In other words, the 
wrapping corrections have to be taken into account at and above the $L^{\text{th}}$ loop 
order for the spin chain of length $L$. There has been many proposals in this direction such 
as thermodynamic Bethe ansatz, Y-system, and quantum spectral curve can be found in literature to 
correctly account for the wrapping corrections
\cite{Sieg:2005kd,Ambjorn:2005wa,Kotikov:2007cy,Gromov:2009tv,Gromov:2013pga}. \par 
Not only the gauge theory but the string theory also witnessed the inefficiency of 
asymptotic Bethe ansatz in the limit of finite volume \cite{SchaferNameki:2005tn}. 
The finite-size corrections to the one-loop calculations in the sigma-model on $AdS_5\times S^5$ has 
been observed as exponentially suppressed in the units of string length 
\cite{SchaferNameki:2006gk,SchaferNameki:2006ey}. The finite-size corrections to giant magnon 
and dyonic giant magnon dispersion relation have been extensively studied at classical 
\cite{Arutyunov:2006gs,Astolfi:2007uz,Klose:2008rx,Ramadanovic:2008qd,Minahan:2008re,
Shenderovich:2008bs,Ahn:2008gd,Jain:2008mt,Hatsuda:2008na,Ahn:2008sk,Grignani:2008te,Bozhilov:2010rf,Floratos:2014gqa} as well as one-loop level \cite{Gromov:2008ie,Heller:2008at}. There have been several instances of finite-size study for certain deformations 
\cite{Bykov:2008bj,Ahn:2010da,Ahn:2014aqa,Ahn:2016egk,Panigrahi:2018xuv,Barik:2019uqf}. Recently, the 
anomalous dimension of GKP strings for finite angular momentum has been 
shown in terms of Lambert $W-$functions in \cite{Floratos:2013cia}. Motivated by all these 
developments, in this paper, we compute finite-size effects for $n$-spiky string using the known 
information of the infinite volume case.\par 
The rest of the paper is organized as follows. In section 2, 
We study the finite-size $n$-spike string in terms of Lambert $\mathbf{W}$-function. We show that for the 
limiting case of $n=2$, it does reduce to the case of the GKP string. Section 3 is devoted to the 
study of finite-size $n$-spike string in $AdS_3 \times S^3$ in the presence of both NS-NS and R-R 
fluxes. In section 4 we study the finite-size $n$ spike strings in pure NS-NS flux. In section 5 we conclude with some discussions.  
\section{Spiky string in AdS}
In this section, we study the finite-size effect on the Kruczenski spiky string in $AdS_3$ background. We start with the $AdS_3$ metric in global coordinates
\begin{eqnarray}\label{metric}
ds^2=- \cosh^2\rho\,dt^2+d\rho^2+\sinh^2\rho\, d\phi^2.
\end{eqnarray}
 To study the relevant string dynamics, we follow closely the analysis presented in \cite{Kruczenski:2004wg}. We choose the following ansatz for the rigidly rotating closed string 
\begin{eqnarray}\label{ansatz}
	t=\tau,~~~ \rho=\rho(\sigma),~~~\phi=\omega \tau+\sigma.
	\end{eqnarray}
	The Nambu-Goto action for the F-string in $AdS_3$ geometry can be written as
	\begin{eqnarray}
	S=-\frac{\sqrt{\lambda}}{2\pi}\int d\tau d\sigma \,\sqrt{-\dot{X}^2 X'^2+(\dot{X}.X')^2}\, ,
	\end{eqnarray}
where $\lambda$ is the 't Hooft coupling constant, $X$'s are the background coordinates and the scalar 
products can be computed using metric (\ref{metric}). The $\dot{X}$ and $X'$ refer to derivative of $X$ with respect to $\tau$ and $\sigma$ respectively. By solving the equation of motion for $\rho$, we get
\begin{eqnarray}\label{eqnrho}
\rho'=\frac{1}{2}\frac{\sinh2\rho}{\sinh2\rho_0}\frac{\sqrt{\sinh^22\rho-\sinh^22\rho_0}}{\sqrt{\cosh^2\rho-\omega^2\sinh^2\rho}}\, ,
	\end{eqnarray}
where $\rho_0$ is the integration constant. We can see from equation (\ref{eqnrho}) that, at 
$\rho=\rho_0$, $\rho'$ vanishes and at $\rho=\coth^{-1}\omega$, $\rho'$ blows up, which indicates the string profile has spikes at maximum value of  $\rho$ (i.e $\rho=\rho_1=\coth^{-1}\omega$) and valleys at the zero of $\rho'$ (i.e 
$\rho=\rho_0$).\\
The angle difference between the spikes and the valleys is given by,
\begin{eqnarray}
\Delta\phi=\theta&=&2\int_{\rho_0}^{\rho_1}\,d\rho \frac{\sinh2\rho_0}{\sinh2\rho}\frac{\sqrt{\cosh^2\rho-\omega^2\sinh^2\rho}}{\sqrt{\sinh^22\rho-\sinh^22\rho_0}}\, .
\end{eqnarray}
Now the energy and angular momentum of the string can be calculated as 
\begin{eqnarray}
E&=&\frac{2n}{2\pi}\frac{\sqrt{\lambda}}{2} \int_{\rho_0}^{\rho_1} d\rho \frac{\sinh\rho}{\cosh\rho}\frac{\left(4 \cosh^4\rho-\omega^2\sinh^22\rho_0\right)}{\sqrt{\left(\cosh^2\rho-\omega^2\sinh^2\rho\right)\left(\sinh^22\rho-\sinh^22\rho_0\right)}}\, ,\\
J&=&\frac{2n}{2\pi}\frac{\sqrt{\lambda}}{2} \omega\int_{\rho_0}^{\rho_1} d\rho \frac{\sinh\rho}{\cosh\rho}\frac{\sqrt{\sinh^22\rho-\sinh^22\rho_0}}{\sqrt{\cosh^2\rho-\omega^2\sinh^2\rho}}\, .
\end{eqnarray}
In energy and angular momentum expression, the integration limit is taken from $\rho_0$ to $\rho_1$ which gives result for one segment arc of closed spiky string. To get the total energy and angular momentum it is multiplied by $2n$.
Now we make the following change of variables 
\begin{eqnarray}\label{variable}
u=\cosh2\rho,~~~~~u_0=\cosh2\rho_0,~~~\text{and}~~~u_1=\cosh2\rho_1.
\end{eqnarray}
Substituting equation (\ref{variable}) in equation (\ref{eqnrho}) and integrating, we get the exact string solution, which has the following form 
\begin{eqnarray}\label{sigma}
\sigma= \frac{\sqrt{u_0^2-1}}{\sqrt{u_1-1}\sqrt{u_0+u_1}}&&\left[\boldmath{\Pi}\left(\arcsin\left(\sqrt{\frac{u_1-u}{u_1-u_0}}\right),\frac{u_1-u_0}{u_1-1},P\right)-\right. \nonumber\\ &&-\left.\boldmath{\Pi}\left(\arcsin\left(\sqrt{\frac{u_1-u}{u_1-u_0}}\right),\frac{u_1-u_0}{u_1+1},P\right)\right],
\end{eqnarray}
where the modulus of complete elliptic integral of third kind $\Pi$ is defined as
\begin{eqnarray}
P=\frac{u_1-u_0}{u_1+u_0}.
\end{eqnarray}
\begin{figure}[H]\label{fig.1}
	\centering
	\includegraphics[scale=.7]{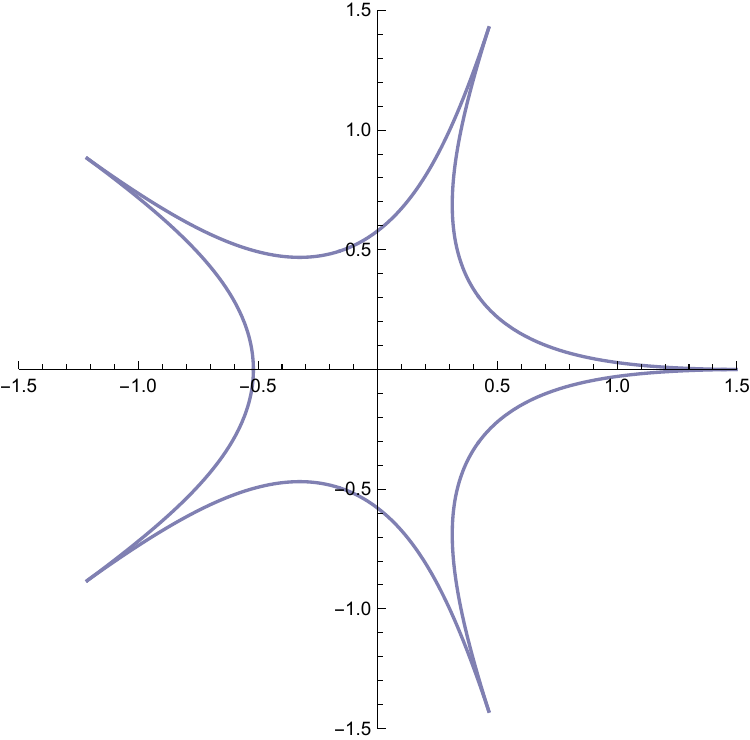}
	\caption{Closed string profile with $n=5$ spikes. Here we have fixed the parameters $\rho_0=0.670835715$ and $\rho_1=1.5$.}
\end{figure}
 The string profile produced by solution (\ref{sigma}) gives a arc segment (half spike) where  $\rho$ goes from $\rho_0$ to  $\rho_1$. Hence the closed string profile can be obtained by gluing $2n$ arc segments. Here, we need to use the closedness condition of the spiky string, which is $\Delta\phi=\frac{2n}{2\pi}$. The plot of closed string profile for $n=5$ spikes in figure 1 has been included for completeness. The change of variables (\ref{variable}) reduces the angular separation, the string energy and the string angular momentum in terms of elliptic integrals as
\begin{eqnarray}
\Delta\phi&=&\theta=\frac{\sqrt{u_0^2-1}}{\sqrt{u_1-1}\sqrt{u_1+u_0}}\left[\boldmath{\Pi}\left(\frac{u_1-u_0}{u_1-1},P\right)-\boldmath{\Pi}\left(\frac{u_1-u_0}{u_1+1},P\right)\right],\\
	E&=&\frac{n\sqrt{\lambda}}{2\pi}\sqrt{u_1-1}\left[\sqrt{u_0+u_1}\boldmath{E}\left(P\right)-\frac{(u_0-1)}{\sqrt{u_0+u_1}}\boldmath{K}\left(P\right)-\right.\nonumber \\
	&&\qquad\qquad\qquad-\left.\frac{u_0^2-1}{(u_1-1)\sqrt{u_0+u_1}}
\boldmath{\Pi}\left(\frac{u_1-u_0}{u_1+1},P\right)\right],\\
	J&=&\frac{n\sqrt{\lambda}}{2\pi}\frac{\sqrt{1+u_1}}{\sqrt{u_0+u_1}}\left[(u_0+u_1)\boldmath{E}\left(P\right)-(1+u_0)\boldmath{K}\left(P\right)-\frac{u_0^2-1}{u_1+1}\boldmath{\Pi}\left(\frac{u_1-u_0}{u_1+1},P\right)\right].\nonumber\\
	\end{eqnarray}
	
Now, we proceed with the following substitution,
\begin{eqnarray}\label{x}
	P=\frac{u_1-u_0}{u_1+u_0}=1-x.
\end{eqnarray}
\begin{figure}[H]\label{fig.1}
	\centering
	\includegraphics[scale=1]{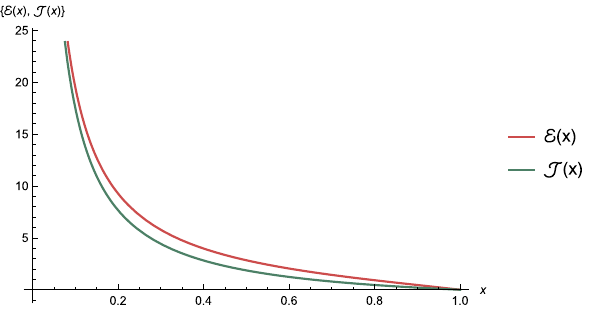}
	\caption{The energy $\mathcal{E}(x)$ and angular momentum  $\mathcal{J}(x)$ of the spiky string for fixed
		$u_0$.}
\end{figure}
With the above substitution and using the formula appendix (\ref{Pi}) for the complete elliptic integral of third kind, the angle difference between the spike and the valley can be written as, 
\begin{eqnarray}\label{phi}
\Delta\phi=\theta&=&\frac{\pi}{2}\frac{1}{\boldmath{K}(x)}\left[\boldmath{F}\left(\arcsin\sqrt{\frac{u_0+1}{2u_0}},x\right)-\frac{\sqrt{(2-x)u_0+x}}{\sqrt{(2-x)u_0-x}}\boldmath{F}\left(\arcsin\sqrt{\frac{u_0-1}{2u_0}},x\right)\right]+\nonumber\\
&+&\frac{1}{\sqrt{2u_0}}\frac{\boldmath{K}(1-x)}{\boldmath{K}(x)}\left[\sqrt{\frac{u_0-1}{1+u_0}}\sqrt{(2-x)u_0-x}\left(\boldmath{\Pi}\left(\frac{1+u_0}{2u_0}x,x\right)-\boldmath{K}(x)\right)-\right.\nonumber\\
&-&\left.\sqrt{\frac{u_0+1}{u_0-1}}\,\frac{\big(\left(2-x\right)u_0+x\big)}{\sqrt{(2-x)u_0-x}}\left(\boldmath{\Pi}\left(\frac{u_0-1}{2u_0}x,x\right)-\boldmath{K}(x)\right)\right].
\end{eqnarray}
Correspondingly, the $(\mathcal{E}-\mathcal{J})$ can be expressed as fuction of $x$ and $u_0$ as
\begin{eqnarray}\label{scalingrelation1}
\mathcal{E}-\mathcal{J}&=&\frac{1}{2}\Bigg[\frac{\boldmath{K}(1-x)}{\sqrt{2u_0}}\left( (1-u_0)\sqrt{(2-x)u_0-x}+(1+u_0)\sqrt{(2-x)u_0+x}\right)+ \nonumber \\
&+&\frac{\sqrt{2u_0}}{x}\left( 
\sqrt{(2-x)u_0-x}-\sqrt{(2-x)u_0+x}\right)\boldmath{E}(1-x)+\frac{(1+u_0)}{\sqrt{2u_0}\boldmath{K}(x)}
\times  \nonumber\\
&\times&\left(\frac{(2-x)u_0+x}{\sqrt{(2-x)u_0-x}}-\sqrt{(2-x)u_0+x}\right)\Bigg(\frac{\pi}{2}
\sqrt{\frac{2u_0(u_0-1)}{(1+u_0)[(2-x)u_0+x]}}\times  \nonumber\\
&&\times\boldmath{F}\left(\arcsin\sqrt{\frac{u_0-1}{2u_0}},x\right)+\boldmath{K}(1-x)\bigg(\boldmath{K
}(x)-\boldmath{\Pi}\left(\frac{u_0-1}{2u_0}x,x\right)\bigg)\Bigg)\Bigg].\nonumber\\
	\end{eqnarray}
Here the energy and the angular momentum of the string have been scaled as $\mathcal{E}\rightarrow 
\frac{\pi}{n\sqrt{\lambda}}E$ and $\mathcal{J}\rightarrow \frac{\pi}{n\sqrt{\lambda}}J$. In figure 2, 
we plot the string energy and angular momentum against the parameter $x$. It is clear from the 
expression (\ref{scalingrelation1}) that to get the string dispersion relation 
$\mathcal{E}(\theta,\mathcal{J})$, we have to first express the parameters $u_0$ and $x$ in terms of 
angle differnce $\Delta\phi=\theta$ and string angular momentum $\mathcal{J}$. For $x \rightarrow 0$, 
the expressions (\ref{phi}) and (\ref{scalingrelation1})  contain logarithmic singularity due to the 
presence of following elliptic functions
	\begin{eqnarray}
	\boldmath{K}(1-x)=\sum_{n=0}^{\infty}(d_n \ln x+h_n) x^n,\\
	 \boldmath{K}(1-x)- \boldmath{E}(1-x)=\sum_{n=0}^{\infty}(c_n \ln x+b_n) x^n.
	\end{eqnarray}
	Here the coefficients that appear in the above series are given by
	\begin{eqnarray}
	d_n=-\frac{1}{2}\left(\frac{(2n-1)!!}{(2n)!!}\right)^2,~~~h_n=-4d_n(\ln2+H_n-H_{2n}),\nonumber\\
	c_n=-\frac{d_n}{2n-1},~~~b_n=-4c_n\left(\ln2+H_n-H_{2n}+\frac{1}{2(2n-1)}\right),
	\end{eqnarray}
	where~~ $H_n=\sum_{k=1}^{n}\frac{1}{k}$.
	The logarithms from the above mentioned expressions can be eliminated using the following expression for $\ln x$, which is obtained using the substitution (\ref{x}) and the elliptic integral $\Pi$ formula (\ref{Pi}) to the angular momentum expression 
	\begin{eqnarray}\label{logx}
	&&\sqrt{\frac{u_0}{(2-x)u_0+x}}2\sqrt{2}\mathcal{J}x+\frac{\pi}{2}\frac{(1+u_0)x}{\boldmath{K}(x)}\,\sqrt{\frac{2u_0(u_0-1)}{(1+u_0)[(2-x)u_0+x]}}\boldmath{F}\left(\arcsin\sqrt{\frac{u_0-1}{2u_0}},x\right)+\nonumber\\
	&&+2u_0\sum_{n=0}^{\infty}b_n x^n-\left(2u_0-\frac{(1+u_0)x}{\boldmath{K}(x)}\,\boldmath{\Pi}\left(\frac{u_0-1}{2u_0}x,x\right)\right)\sum_{n=0}^{\infty}h_n x^n=\ln x\left[-2u_0\sum_{n=0}^{\infty}c_n x^n+\right.\nonumber\\
	&&+\left.\left(2u_0-\frac{(1+u_0)x}{\boldmath{K}(x)}\,\boldmath{\Pi}\left(\frac{u_0-1}{2u_0}x,x\right)\right)\sum_{n=0}^{\infty}d_n x^n\right].
	\end{eqnarray}
	Setting $u=\csc a$, the equation (\ref{phi}) can be expanded in a double series around both $x=0$ and $a=\theta$, then the series can be inverted for $a$ by using Mathematica. Now plugging the expression $u_0=\csc a=u_0(x,\theta,\mathcal{J})$ into (\ref{logx}) and exponentiating, we end up with the following equation 
	\begin{eqnarray}\label{x1}
	x=x_0\, \exp\left(\frac{a_0}{x}+a_1 x+a_2x^2+...\right),
	\end{eqnarray}
	where the coefficients $a_n=a_n(\theta,\mathcal{J})$ can be obtained from the equation (\ref{logx}) and $x_0$ is given by
	 \begin{eqnarray}
	x_0=16\, \exp\left[4\mathcal{J}-1+\left(\pi-2\theta\right)\cot\theta+\frac{\cot^2\theta}{2}+\csc\theta\left(2+\frac{\csc\theta}{2}\right)\right]=16\, \exp(4\mathcal{J}+\mathcal{A}).\nonumber\\
	\end{eqnarray}
The expression  $x(\theta,\mathcal{J})$ can be obtained from equation (\ref{x1}). But we can see, it contains $1/x$ term in the exponent and also $x_0$ is exponentially increasing with $\mathcal{J}$. So the general Lagrange-B$\ddot{\text{u}}$rmann formula cannot be applied to get $x$, rather a different approach has to be adopted. Following the procedure discussed in \cite{Floratos:2013cia}, let us define $x^*$ as the leading solution to (\ref{x1})	
\begin{eqnarray}
x^*=x_0~ e^{a_0/x^*},
\end{eqnarray}  
which implies the functional form of $x^*$ as 
\begin{eqnarray}\label{x*}
x^*=\frac{a_0}{\mathbf{W}_{-1}\big(a_0/x_0\big)}=\frac{-4\csc\theta}{\mathbf{W}_{-1}\Big(-\frac{\csc\theta}{4}\exp\left(-4\mathcal{J}-\mathcal{A}\right)\Big)}\,.
\end{eqnarray}
 For $\mathcal{J}\rightarrow\infty$, $x^*$ approaches to zero $(x^*\rightarrow 0)$. Hence, we have chosen $\mathbf{W}_{-1}$, the lower branch of Lambert function over $\mathbf{W}_{0}$, the principal branch\footnote{The equation (\ref{x1}) has to be solved in limit $x\rightarrow 0$ and $\mathcal{J}\rightarrow\infty$. We can see, in this limit $x_0\rightarrow\infty$ and the coefficient $a_0$, which can be computed from the series expansion of (\ref{logx}) is found to be $a_0<0$. So in this region, for $\mathcal{J}\rightarrow\infty$, $\mathbf{W}_{0}\rightarrow 0$ and $\mathbf{W}_{-1}\rightarrow -\infty$. If we choose $\mathbf{W}_{0}$ instead of $\mathbf{W}_{-1}$, $x^*$ will blows up which is inappropriate for our case.} with argument $a_0/x_0=-\frac{1}{4}\csc\theta\exp(-4\mathcal{J}-\mathcal{A})$. Considering $x^*$ as the leading solution to (\ref{x1}), we can set the full solution 
	  \begin{eqnarray}
	  x=x^* e^z, \quad\text{with}\,\, z\rightarrow0.
	  \end{eqnarray}
	  Substituting the above equation into (\ref{x1}) and using (\ref{x*}),
	  we get the following expression for $z$ 
	  \begin{eqnarray}
	  z=\frac{a_1}{a_0}(x^*)^2+\left(\frac{a_2}{a_0}-\frac{a_1}{a_0^2}\right)(x^*)^3+\left(\frac{a_1}{a_0^3}+\frac{3a_1^2-2a_2}{2a_0^2}+\frac{a_3}{a_0}\right)(x^*)^4+...
	  \end{eqnarray}
	   which leads to the $x$ as a series in $x^*$
	   \begin{eqnarray}
	   x=x^*+\frac{a_1}{a_0}(x^*)^3+\left(\frac{a_2}{a_0}-\frac{a_1}{a_0^2}\right)(x^*)^4+\left(\frac{a_1}{a_0^3}+\frac{2a_1^2-a_2}{a_0^2}+\frac{a_3}{a_0}\right)(x^*)^5+...
	   \end{eqnarray}
	    Now substituting $u_0\left(\mathcal{J},\theta\right)$ and $x\left(\mathcal{J},\theta\right)$ to  $(\mathcal{E}-\mathcal{J})$ in equation (\ref{scalingrelation1}),  we get the desired finite-size dispersion relation for closed rigidly rotating string in terms of Lambert function  $\mathbf{W}_{-1}$ with argument $(-\frac{1}{4}\csc\theta\exp(-4\mathcal{J}-\mathcal{A}))$,
	   \begin{eqnarray}\label{scalingrelation2}
	   \mathcal{E}-\mathcal{J}&=&-\frac{1}{2}\mathbf{W}_{-1}-2\mathcal{J}-\frac{1}{2}\left(\pi-2\theta\right)\cot\theta-\frac{1}{8}\csc^2\theta\big(3+\cos2\theta+8 \sin\theta\big)-\nonumber\\
	   &&-\frac{1}{8}\Big(2-8\csc\theta-3\csc^2\theta-2\left(\pi-2\theta\right)\cot\theta\Big)\frac{1}{\mathbf{W}_{-1}}+...
	   \end{eqnarray} 
	   To see the explicit finite-size corrected structure, we would like to calculate the leading order correction terms (leading in $\mathcal{J}$) to the infinite-size dispersion relation by expanding the Lambert $\mathbf{W}_{-1}$ function for large value of angular momentum $\mathcal{J}$. For this purpose we have taken the following few terms, which are expected to contribute to the leading terms of the scaling relation,
	   \begin{eqnarray}\label{leadingDR}
	   \left.\mathcal{E}-\mathcal{J}\right\vert_{(\text{L+...})}=\frac{2}{x^*}\csc\theta-2\mathcal{J}-\left(\frac{\pi}{2}-\theta\right)\cot\theta-\frac{1}{8}\csc^2\theta\left(3+\cos2\theta+8 \sin\theta\right). \nonumber\\
	   \end{eqnarray}
	   Using the expansion of the $\mathbf{W}_{-1}$ function, we write the explicit form of $1/x^*$ as
	  \begin{eqnarray}{\label{x**}}
	  \frac{1}{x^*}&=&\frac{1}{\csc\theta}\left[\mathcal{J}+\frac{\ln \mathcal{J}}{4}+\ln 2+\frac{\ln(\sin\theta)}{4}+\frac{\mathcal{A}}{4}+\sum_{k=1}^{\infty}\frac{(-1)^{k-1}}{4k}\left(\frac{\ln(4\sin\theta)+\mathcal{A}}{4\mathcal{J}}\right)^k-\right.\nonumber\\
	  &-&\left.\sum_{i,n=0}^{\infty}\,\sum_{m=1}^{\infty}\,\sum_{j=0}^{m}\,\frac{(-1)^m}{4^{n+m+1}m!}\left[\begin{array}{cc} n+m \\
	 n+1 \\ \end{array}\right]\left(\begin{array}{cc}-n-m \\
	 i \\ \end{array}\right)\left(\begin{array}{cc} m \\
	 j \\ \end{array}\right)\frac{(\ln\mathcal{J})^j}{\mathcal{J}^{n+m}}\times\right.\nonumber\\
	 &\times& \left. \left(\frac{\ln(4\sin\theta)+\mathcal{A}}{4\mathcal{J}}\right)^i\left(2\ln 2+\sum_{k=1}^{\infty}\frac{(-1)^{k-1}}{k}\left(\frac{\ln(4\sin\theta)+\mathcal{A}}{4\mathcal{J}}\right)^k\right)^{m-j}\right].
	  \end{eqnarray}
	 The above expression have been written by considering $\mathcal{J}$ large and we discuss the Stirling numbers of first kind $\left[\begin{array}{cc} n+m \\
	 n+1 \\ \end{array}\right]$  in appendix A. Plugging (\ref{x**}) to (\ref{leadingDR}), we obtain the following expression for the leading order correction terms
\begin{eqnarray}\label{DL1}
\left.\mathcal{E}-\mathcal{J}\right\vert_{(\text{L+...})}&=&\frac{\ln \mathcal{J}}{2}+2\ln 2+\frac{\ln(\sin\theta)}{2}-\frac{1}{2}+\sum_{k=1}^{\infty}\frac{(-1)^{k-1}}{2k}\left(\frac{\ln(4\sin\theta)+\mathcal{A}}{4\mathcal{J}}\right)^k-\nonumber\\
&&-\frac{1}{2}\sum_{i,n=0}^{\infty}\sum_{m=1}^{\infty}\sum_{j=0}^{m}\frac{(-1)^m}{4^{n+m}m!}\left[\begin{array}{cc} n+m \\
n+1 \\ \end{array}\right]\left(\begin{array}{cc}-n-m \\
i \\ \end{array}\right)\left(\begin{array}{cc} m \\
j \\ \end{array}\right)\frac{(\ln\mathcal{J})^j}{\mathcal{J}^{n+m}}\times\nonumber\\
&&\times \left(\frac{\ln(4\sin\theta)+\mathcal{A}}{4\mathcal{J}}\right)^i\left(2\ln 2+\sum_{k=1}^{\infty}\frac{(-1)^{k-1}}{k}\left(\frac{\ln(4\sin\theta)+\mathcal{A}}{4\mathcal{J}}\right)^k\right)^{m-j}.\nonumber\\
\end{eqnarray}
One may notice, the finite-size dispersion relation for spiky string reproduces the result of infinite string, which has been discussed in \cite{Kruczenski:2004wg} with the correction terms. For $n=2$ (closedness condition implies $\Delta\phi=\theta=\pi/2$), our results (\ref{scalingrelation2}) and (\ref{DL1}) are in complete agreement with the results of finite-size correction to the closed folded GKP string dissused in \cite{Floratos:2013cia}. 
\section{Spiky string with mixed fluxes}
In this section, we discuss the finite-size effect on the spiky string in $AdS_3$ background supported by mixed NS-NS and R-R fields
\begin{eqnarray}
B_{t\phi}=b\sinh^2\rho, \quad C_{t\phi}=\sqrt{1-b^2}\sinh^2\rho,
\end{eqnarray}
where the parameter $b$ varies from $0$ to $1$ $(0\le b\le 1)$. Since we are dealing with fundamental string solution, the R-R field is not relevant to us and can be ignored.
\begin{figure}[H]\label{fig.2b}
	\centering
	\includegraphics[scale=.6]{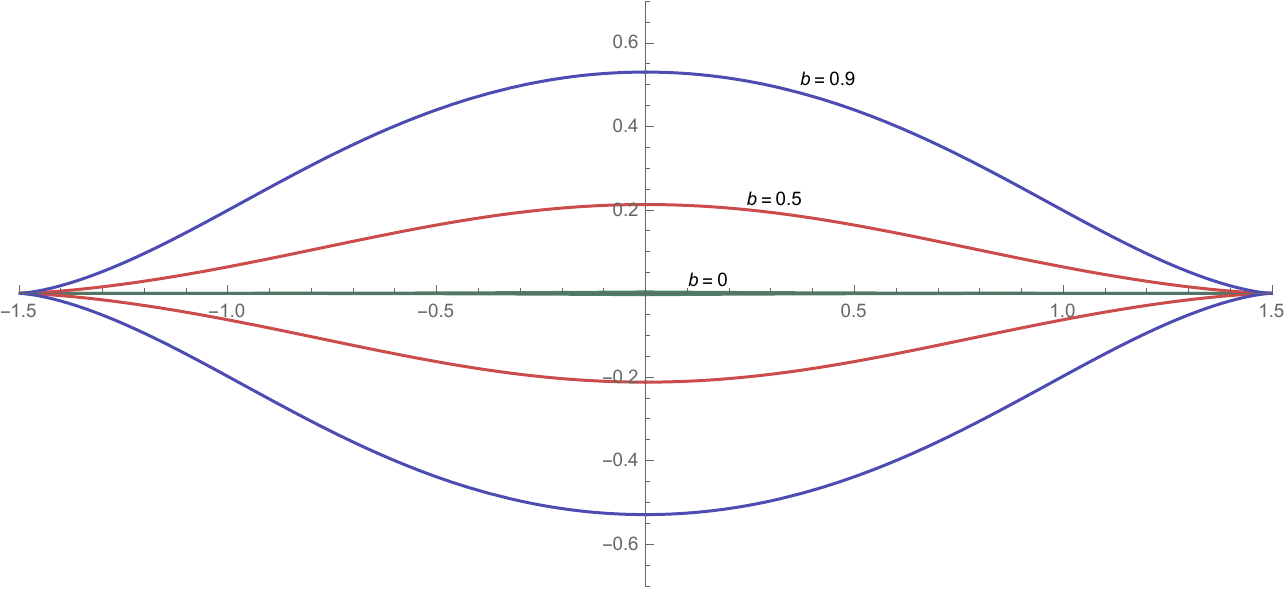}
	\caption{Closed string profile with $n=2$ spikes in $AdS$ space with different values of the flux parameter $b$. Here, we have fixed the spike position at $\rho_1=1.5$.}
\end{figure}
The Nambu-Goto action for the F-string in presence of the mixed fluxes is written as 
\begin{eqnarray}
S=-\frac{\sqrt{\lambda}}{2\pi}\int d\tau d\sigma \,\left[\sqrt{-\dot{X}^2 X'^2+(\dot{X}.X')^2}-\frac{\epsilon^{ab}}{2}B_{ij}\partial_aX^i\partial_bX^j\right],
\end{eqnarray}
where $\lambda$ is the 't Hooft coupling as mentioned in the previous section and $\epsilon^{ab}$ is the antisymmetric tensor with $\epsilon^{\tau\sigma}=-\epsilon^{\sigma\tau}=1$. We use the same ansatz as equation (\ref{ansatz}). The equations of motion for $t$ and $\theta$ are satisfied if
	\begin{eqnarray}\label{eqnb}
\rho'=\frac{\sinh2\rho}{4\left(\mathcal{K}+b\sinh^2\rho\right)}\frac{\sqrt{\sinh^22\rho-4\left(\mathcal{K}+b\sinh^2\rho\right)^2}}{\sqrt{\cosh^2\rho-\omega^2\sinh^2\rho}}\, ,
\end{eqnarray}
which is consistent with the equation of motion for $\rho$. Here $\mathcal{K}$ in the above expression is an integration constant. It can be noticed that the presence of B-field does not change the maximum value of $\rho$ ($\rho_{1}=\coth^{-1}\omega$), the cusp position but the minimum value has been modified by the parameter $b$. Now we have two roots $(\rho_0^{\pm})$ corresponding to the zero of $\rho'$, the valley positon but we will consider the one which reduces to $\rho_0$ for $b=0$. Accordingly the conserved charges and angle difference between the spike and the valley have also been changed due to inclusion of mixed flux.
\begin{figure}[H]\label{fig.3b}
	\centering
	\includegraphics[scale=.7]{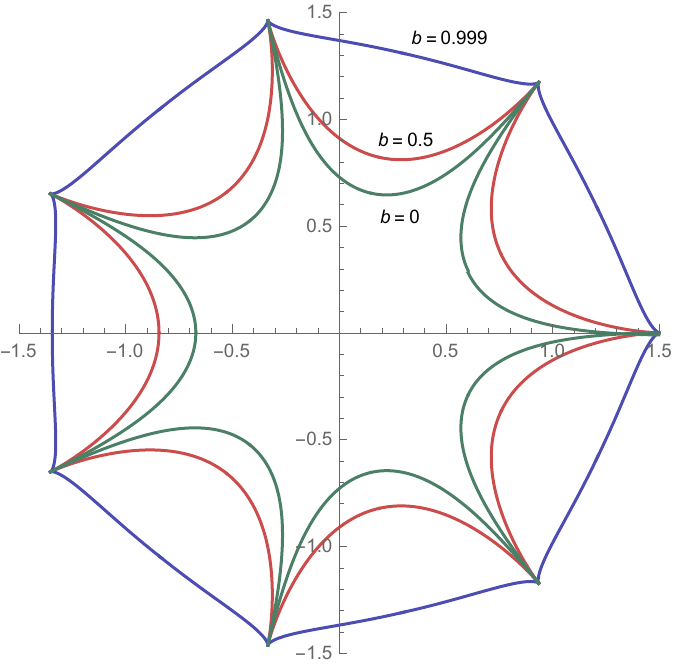}
	\caption{Closed spiky string profile with $n=7$ spikes in $AdS$ space with different values of the flux parameter $b$. Here, the cusp position has been fixed at $\rho_1=1.5$.}
\end{figure}
The angle difference between the spike and the valley of one arc segment of closed spiky string is given by
\begin{eqnarray}
\Delta \phi=\theta=4 \int_{\rho_0}^{\rho_1} d\rho \frac{\left(\mathcal{K}+b\sinh^2\rho\right)}{\sinh2\rho}\frac{\sqrt{\cosh^2\rho-\omega^2\sinh^2\rho}}{\sqrt{\sinh^22\rho-4\left(\mathcal{K}+b\sinh^2\rho\right)^2}}.
\end{eqnarray} 
The energy and the angular momentum of the string are obtained as follows
\begin{eqnarray}
E&=&\frac{2n\sqrt{\lambda}}{\pi} \int_{\rho_0}^{\rho_1} d\rho\, \tanh\rho \left[\frac{\left( \cosh^4\rho-\omega^2\left(\mathcal{K}+b\sinh^2\rho\right)^2\right)}{\sqrt{\left(\cosh^2\rho-\omega^2\sinh^2\rho\right)\left(\sinh^22\rho-4\left(\mathcal{K}+b\sinh^2\rho\right)^2\right)}}-\right.\nonumber \\
&&\qquad\qquad\qquad\qquad\qquad\qquad-\left.b\frac{\left(\mathcal{K}+b\sinh^2\rho\right)\sqrt{\left(\cosh^2\rho-\omega^2\sinh^2\rho\right)}}{\sqrt{\left(\sinh^22\rho-4\left(\mathcal{K}+b\sinh^2\rho\right)^2\right)}}\right],
\end{eqnarray} 
\begin{eqnarray}
J=\frac{n\sqrt{\lambda}}{2\pi} \omega\int_{\rho_0}^{\rho_1} d\rho\,\tanh\rho\frac{\sqrt{\sinh^22\rho-4\left(\mathcal{K}+b\sinh^2\rho\right)^2}}{\sqrt{\cosh^2\rho-\omega^2\sinh^2\rho}}. 
\end{eqnarray}
Here $n$ is the number of spikes and using the string closedness condition, we can express $\Delta\phi$ in terms of $n$ as  $\Delta\phi=\frac{2\pi}{2n}$.
\newpage 
Let's proceed by making the following change of variable like the previous section 
\begin{eqnarray}\label{variableb}
u=\cosh2\rho,~~~~~u_0^\pm=\cosh2\rho_0^\pm,~~~\text{and}~~~u_1=\cosh2\rho_1\, .
\end{eqnarray}
The two roots corresponding to the valley of the string configuration can be calculated as, 
\begin{eqnarray}
u_0^\pm=\frac{-\left(b^2-2b\mathcal{K}\right)\pm\sqrt{1+4\mathcal{K}^2-4b\mathcal{K}}}{1-b^2}\, .
\end{eqnarray}
Clearly $u_0^+$ is the correct one, we are looking for as it return to the original root $u_0$ for $b=0$.
Using the above substitutions to the equation (\ref{eqnb}) and performing the integration, the exact string solution has been obtained in the following form
\begin{eqnarray}\label{sigmab}
\sigma&=& \frac{2}{\sqrt{1-b^2}\sqrt{\left(u_1-1\right)\left(u_1-u_0^-\right)}}\left[\mathcal{K}\,\boldmath{\Pi}\left(\frac{u_1-u_0^+}{u_1-1},\alpha,P\right)-\right. \nonumber \\
&&\left.\qquad\qquad\qquad\qquad\quad-\left(\mathcal{K}-b\right)\boldmath{\Pi}\left(\frac{u_1-u_0^+}{u_1+1},\alpha,P\right)-b\,\boldmath{F}\left(\alpha,P\right)\right], 
\end{eqnarray}
where the arguments of elliptic function are written as
\begin{eqnarray}
\sin\alpha=\sqrt{\frac{u_1-u}{u_1-u_0^+}},  \quad\text{and}\quad P=\frac{u_1-u_0^+}{u_1-u_0^-}.
\end{eqnarray}
Again for the purpose of completeness and to make a clear understanding of the impact of flux on string profile, we have plotted closed string profiles in figure 3 and figure 4, for $n=2$ spikes and $n=7$ spikes respectively considering different values of flux parameter $b$. It can be observed that the inclusion of B-field makes the string profile more fatter and the string becomes almost circular when $b\rightarrow1$.
Using the substitution of variables (\ref{variableb}), the anglular separation and the conserved charges can be expressed in terms of complete elliptic integrals as follows
\begin{eqnarray}\label{phib}
\Delta\phi=\theta&=&\frac{2}{\sqrt{1-b^2}\sqrt{\left(u_1-1\right)\left(u_1-u_0^-\right)}}\left[\mathcal{K}\,\boldmath{\Pi}\left(\frac{u_1-u_0^+}{u_1-1},P\right)-\right.\nonumber\\
&&\left.\qquad\qquad\qquad\qquad\quad-\left(\mathcal{K}-b\right)\boldmath{\Pi}\left(\frac{u_1-u_0^+}{u_1+1},P\right)-b\,\boldmath{K}\left(P\right)\right],
\end{eqnarray}
\begin{eqnarray}\label{eb}
E&=&\frac{n\sqrt{\lambda}}{2\pi}\frac{1}{\sqrt{(1-b^2)(u_1-1)}}\left[\frac{1}{\sqrt{u_1-u_0^-}}\left((1+b^2)(u_1-1)-4b\mathcal{K}u_1\right)\boldmath{K}(P)+\right. \nonumber\\
&&\left.\qquad\qquad+\,(1-b^2)(u_1-1)\left(\frac{u_0^-}{\sqrt{u_1-u_0^-}}\boldmath{K}(P)+\sqrt{u_1-u_0^-}\boldmath{E}(P)\right)+\right.\nonumber\\
&&\qquad\qquad\qquad\qquad\qquad\qquad+\left.\frac{4\mathcal{K}\left(b-\mathcal{K}\right)}{\sqrt{u_1-u_0^-}}\boldmath{\Pi}\left(\frac{u_1-u_0^+}{u_1+1},P\right)\right],
\end{eqnarray}
\begin{figure}[H]
	\centering
	\begin{subfigure}[t]{0.5\textwidth}
		\centering
		\includegraphics[width=1\textwidth]{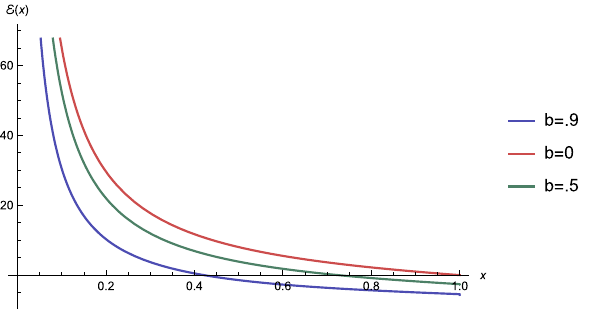}
		\caption{}
	\end{subfigure}%
	~ 
	\begin{subfigure}[t]{0.6\textwidth}
		\centering
		\includegraphics[width=1\textwidth]{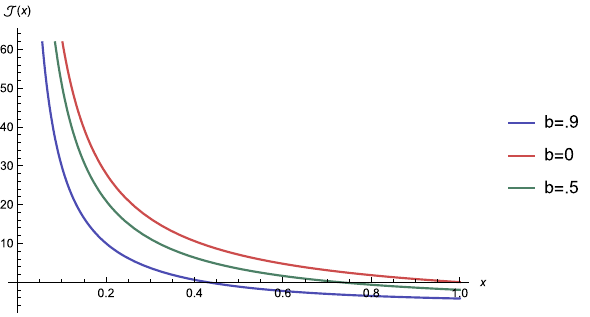}
		\caption{}
	\end{subfigure} 
	
	\caption{ The energy $\mathcal{E}(x)$ and angular momentum  $\mathcal{J}(x)$ of the rotating string for different values of flux parameter $b$ with fixed $u_0$.}
\end{figure}
\begin{eqnarray}\label{jb}
J&=&\frac{n\sqrt{\lambda}}{2\pi}\sqrt{\frac{1+u_1}{1-b^2}}\left[(1-b^2)\left(\frac{u_0^-}{\sqrt{u_1-u_0^-}}\boldmath{K}(P)+\sqrt{u_1-u_0^-}\boldmath{E}(P)\right)+\right. \nonumber\\
&&\quad+\left.\frac{\left(3b^2-4b\mathcal{K}-1\right)}{\sqrt{u_1-u_0^-}}\boldmath{K}(P)-\frac{4\left(\mathcal{K}-b\right)^2}{(1+u_1)\sqrt{u_1-u_0^-}}\boldmath{\Pi}\left(\frac{u_1-u_0^+}{u_1+1},P\right)\right].
\end{eqnarray}
Now we proceed to get the finite-size correction to the spinning string solution with B-field by making the following substitution as before for the argument of elliptic function
\begin{eqnarray}\label{xb}
P=\frac{u_1-u_0^+}{u_1-u_0^-}=1-x.
\end{eqnarray}
The Figure (5a) and (5b) depicts the string energy $\mathcal{E}(x)$ and angular momentum $\mathcal{J}(x)$ for different values of flux parameter $b$ with fixed $u_0$.
It can be noticed that, the expressions for angle of separation between spike and valley, angular momentum and energy of the string are quite complicated in presence of flux. So to make the evaluation easy, first we expand the expressions (\ref{phib})-(\ref{jb}) around small $b$ upto the second order in $b$ (we restrict ourselves to $\mathcal{O}(b^2)$ due to very large and complicated output for $n$ spikes). Now, we have paramere $u_0$ instead of $u_0^+$ in our expressions. Henceforth applying the same procedure we have followed in the last section, we first express $u_0$ as  $u_0\left(\mathcal{J},\theta,x\right)$. 
\newpage
Substituting  $u_0\left(\mathcal{J},\theta,x\right)$ to the angular momentum expression, the following simplified expression has been computed
\begin{eqnarray}\label{xb}
x=x_0\, \exp\left(\frac{a_0}{x}+a_1 x+a_2x^2+...\right),
\end{eqnarray}
where the coefficients $a_n=a_n(\theta,\mathcal{J})$ can be evaluated from $\mathcal{J}$ expression and the $x_0$ has the form
\begin{eqnarray}
x_0&=&16\, \exp\left[4\mathcal{J}-1+\left(\pi-2\theta\right)\cot\theta+\frac{\cot^2\theta}{2}+\csc\theta\left(2+\frac{\csc\theta}{2}\right)+\mathcal{B}\right],\nonumber\\
&=&16\, \exp(4\mathcal{J}+\mathcal{A}+\mathcal{B}).
\end{eqnarray}
where $\mathcal{B}$ can be expressed as
\begin{eqnarray}
\mathcal{B}&=&\bigg(-8\mathcal{J}\cot\theta-\frac{1}{16}\csc^3\theta\Big(70\cos\theta+12\cos3\theta-2\cos5\theta+4(6\sin2\theta+\sin4\theta)+\quad\nonumber\\
&&+(\pi-2\theta)\big(32\sin\theta+7\sin3\theta-\sin5\theta\big)\Big)\bigg)b+\bigg(\left(-10+4\cos2\theta+24\csc^2\theta\right)\mathcal{J}+\nonumber\\
&&+\frac{1}{512}\csc^4\theta\Big(6085+4708\cos2\theta+60\cos4\theta-100\cos6\theta-\cos8\theta+1880\sin\theta+\nonumber\\
&&+952\sin3\theta+88\sin5\theta-8\sin7\theta+(\pi-2\theta)\big(2454\sin2\theta+130\sin4\theta-\nonumber\\
&&-50\sin6\theta-\sin8\theta\big)\Big)\bigg)b^2.
\end{eqnarray}
Now we can write the leading solution to (\ref{xb}) as
\begin{eqnarray}
x^*&=&\frac{a_0}{\mathbf{W}_{-1}\big(a_0/x_0\big)},
\end{eqnarray}
using which the full solution for $x(\theta,\mathcal{J})$ can be constructed (already discussed in the last section). It should be mentioned that, here the coefficient $a_0(\theta,\mathcal{J})$
has been modified to
\begin{eqnarray}
a_0=-4\csc\theta+2\left(3+\cos2\theta\right)\cot\theta\csc\theta\,b-\frac{1}{2}\left(40\cot^2\theta\csc\theta+7\sin\theta-\sin3\theta\right)b^2.\nonumber\\
\end{eqnarray} 
\newpage
Finally plugging the parameters $u_0\left(\mathcal{J},\theta,x\right)$ and $x\left(\mathcal{J},\theta\right)$ to the $\mathcal{E}-\mathcal{J}$ expression, we write the finite-size dispersion relation for closed folded spiky string in presence of mixed fluxes as
\begin{eqnarray}\label{scalingrelationb}
\mathcal{E}-\mathcal{J}&=&\frac{-1}{2}\left(1-\frac{b^2}{2} \right)\mathbf{W}_{-1}-2\mathcal{J}-\frac{1}{2}\left(\pi-2\theta\right)\cot\theta-\frac{1}{8}\csc^2\theta\left(3+\cos2\theta+8 \sin\theta\right)+\nonumber\\
&&+\bigg(4\mathcal{J}\cot\theta+\frac{1}{32}\csc^3\theta\Big(68\cos\theta+13\cos3\theta-\cos5\theta+4(6\sin2\theta+\sin4\theta)+ \nonumber\\
&&+(\pi-2\theta)\big(26\sin\theta+9\sin3\theta-\sin5\theta\big)\Big)\bigg) b+\bigg(2J\big(3-\cos2\theta-6 \csc^2\theta\big) -\nonumber\\
&&-\frac{1}{1024}\csc^4\theta  \Big(5673+5028 \cos2\theta+ 
156 \cos4\theta -100\cos6\theta- 5 \cos8\theta+
\nonumber\\
&&+1496 \sin\theta+1080\sin3\theta +88 \sin5\theta-
8 \sin7\theta +(\pi - 
2 \theta) \big(2390 \sin2\theta +\nonumber\\
&&+162 \sin4\theta-50 \sin6\theta - \sin8\theta\big)\Big)\bigg)b^2-\frac{1}{8}\bigg(2-8\csc\theta-3\csc^2\theta-\nonumber\\
&&-2\left(\pi-2\theta\right)\cot\theta\bigg)\frac{1}{\mathbf{W}_{-1}}-b\bigg(2J\cot\theta +\frac{1}{32}\csc^3\theta
\Big(60\cos\theta+5\cos3\theta-\nonumber\\
&&-\cos5\theta+4\left(6\sin2\theta +\sin4\theta\right) +\left(\pi-2\theta\right)\big(16\sin\theta+7\sin3\theta-\sin5\theta\big)\Big)\bigg)\frac{1}{\mathbf{W}_{-1}}+\nonumber\\
&&+b^2 \bigg(2J \left(-2+\cos2\theta+4\csc^2\theta\right) + \frac{1}{128}\csc^4\theta\Big(667+502\cos2\theta+6\cos4\theta-\nonumber\\
&&-8\cos6\theta +\cos8\theta+ 127\sin\theta + 
135\sin3\theta+ 11\sin5\theta-\sin7\theta+\left(\pi-2\theta\right)\times\nonumber\\
&&\times\big(222 \sin2\theta+16\sin4\theta- 6\sin6\theta+\sin8\theta\big)\Big)\bigg)\frac{1}{\mathbf{W}_{-1}}+...
\end{eqnarray}	
The above scaling relation agrees well with the infinite-size result of spiky string with flux discussed in \cite{Banerjee:2015qeq}. One may notice that turning B-field off, dispersion relation (\ref{scalingrelationb}) gives us back the finite-size $\left(\mathcal{E}-\mathcal{J}\right)$ relation (\ref{scalingrelation2}). For $n=2$, we get the following dispersion relation in presence of mixed fluxes
\begin{eqnarray}
\mathcal{E}-\mathcal{J}&=&\frac{-1}{2}\left(1-\frac{b^2}{2} \right)\mathbf{W}_{-1}-2\mathcal{J}-\frac{5}{4}-\left(4\mathcal{J}+\frac{11}{8}\right)b^2+\nonumber\\
&&\qquad+\left(\frac{9}{8}+\left(2\mathcal{J}+\frac{23}{16}\right)b^2\right)\frac{1}{\mathbf{W}_{-1}}+...\,\, ,\nonumber\\
\end{eqnarray}
where the argument of $\mathbf{W}$-function is $\left(-\frac{1}{4}(1+b^2)\exp\left[-4\mathcal{J}-3/2-\left(10\mathcal{J}+5\right)b^2\right]\right)$.
\section{Spiky string with pure NS-NS flux} 
It would be interesting to discuss the special case, where the $AdS$ background supported merely by NS-NS flux. This instance can be achieved by taking the flux parameter $b=1$, where the R-R flux vanishes. The substitution of $b=1$ reduces one constant parameter from the theory which makes the problem more tractable.\par
In this case, the equation of motion for $\rho$ comes out to be
\begin{eqnarray}\label{eqnb1}
\rho'=\frac{\sinh2\rho}{4\left(\mathcal{K}+\sinh^2\rho\right)}\frac{\sqrt{\sinh^22\rho-4\left(\mathcal{K}+\sinh^2\rho\right)^2}}{\sqrt{\cosh^2\rho-\omega^2\sinh^2\rho}},
\end{eqnarray}
which is simply the $b=1$ limit of the equation (\ref{eqnb}). Like the previous case of mixed flux, here also the spike position remains same as $\rho_1=\coth^{-1}\omega$, where as the position of valley $\rho_0$ has been modified.
Using the following change of variables
\begin{eqnarray}
u=\cosh2\rho,~~~~~u_0=\cosh2\rho_0,~~~\text{and}~~~u_1=\cosh2\rho_1\, ,
\end{eqnarray} 
 we obtain the exact string profile in terms of trigonometric functions as 
\begin{eqnarray}
\sigma&=&\frac{1}{\sqrt{2(u_1-1)\left(u_0-\sqrt{u_0^2-1}\right)}}\Bigg[\tan^{-1}\left(\frac{u_1+u_0-2u}{2\sqrt{u-u_0}\sqrt{u_1-u}}\right)+\nonumber\\
&&\qquad+\,\frac{\sqrt{u_1-1}\left(u_0-\sqrt{u_0^2-1}-1\right)}{\sqrt{u_0-1}}\tan^{-1}\left(\frac{\sqrt{u_0-1}\sqrt{u_1-u}}{\sqrt{u_1-1}\sqrt{u-u_0}}\right)-\nonumber\\
&&\qquad-\,\frac{\sqrt{u_1+1}\left(u_0-\sqrt{u_0^2-1}+1\right)}{\sqrt{u_0+1}}\tan^{-1}\left(\frac{\sqrt{u_0+1}\sqrt{u_1-u}}{\sqrt{u_1+1}\sqrt{u-u_0}}\right)\Bigg].
\end{eqnarray}
The explicit form of $u_0$ in terms of integration constant $\mathcal{K}$ can be written as
\begin{eqnarray}
u_0=\frac{2\mathcal{K}^2-2\mathcal{K}+1}{1-2\mathcal{K}}.
\end{eqnarray}
The angular separation $\Delta\phi$ between the valley $(\rho=\rho_0)$ and the spike $(\rho=\rho_1)$ in the limit $\rho_1>>\rho_0>>1$ takes the following form
\begin{eqnarray}
\Delta\phi&=&\frac{-\pi}{2\sqrt{2}\sqrt{(u_1-1)\left(u_0-\sqrt{u_0^2-1}\right)}}\Bigg[2+\left(u_0-\sqrt{u_0^2-1}-1\right)\frac{\sqrt{u_1-1}}{\sqrt{u_0-1}}-\nonumber\\
&&\qquad\qquad\qquad\qquad\qquad\qquad\quad-\left(u_0-\sqrt{u_0^2-1}+1\right)\frac{\sqrt{u_1+1}}{\sqrt{u_0+1}}\,\Bigg].
\end{eqnarray}
The string angular momentum and energy expressions in the above mentioned limit can be evaluated as follows
\begin{eqnarray}
J=\frac{n\sqrt{\lambda}}{\pi}\mathcal{J}=\frac{n\sqrt{\lambda}}{2\sqrt{2}}\,\sqrt{u_0-\sqrt{u_0^2-1}}\,\Big[\sqrt{u_1+1}-\sqrt{u_0+1}\Big],
\end{eqnarray}
\begin{eqnarray}
E&=&\frac{n\sqrt{\lambda}}{\pi}\mathcal{E}=\frac{n\sqrt{\lambda}}{2\sqrt{2}\sqrt{(u_1-1)\left(u_0-\sqrt{u_0^2-1}\right)}}\left[\left(1-u_0^2+u_0\sqrt{u_0^2-1}\right)\times\right.\nonumber\\
 &&\qquad\qquad\qquad\qquad\qquad\times\left.\frac{\sqrt{u_1+1}}{\sqrt{u_0+1}}-
\left(1-u_1\left(u_0-\sqrt{u_0^2-1}\right)\right)\right].
\end{eqnarray}
Finally, the dispersion of closed rotated string with pure NS-NS flux has the following form
\begin{eqnarray}
\mathcal{E}-\mathcal{J}=\frac{\pi}{2}-\frac{\pi^2}{8\mathcal{J}}+\frac{\pi^3}{32\mathcal{J}^2}-\frac{\pi^4}{128\mathcal{J}^3}+\frac{\pi^5}{512\mathcal{J}^4}-\frac{\pi^6}{2048\mathcal{J}^5}+ \cdots
\end{eqnarray}
Note that for $\mathcal{J}\rightarrow\infty$, the scaling relation reduces to $\mathcal{E}-\mathcal{J} = \pi/2$, which is the analogous relation in the infinite charge limit.\footnote{Interestingly, the above series reduces to a geometric series of the form $\mathcal{E}-\mathcal{J}=\frac{2\pi \mathcal{J}}{\pi+4\mathcal{J}} .$}
\section{Conclusions}
In this paper, we have studied finite-size spiky string in $AdS_3\times S^3$ background with and without flux. Firstly, we have discussed the finite-size dispersion relation for the spiky string in $AdS_3$ background in terms of Lambert $\mathbf{W}$-function. When the background is supported by mixed NS-NS and R-R fluxes, we have followed the perturbation technique to analyse the associated string dynamics. The scaling relation between the energy and the angular momentum has been computed for general $n$ in the presence of mixed fluxes. We have further discussed the limiting $n=2$ case in detail, which resembles to the long folded string. For pure NS-NS case, the result has been presented by naive geometric series. The analysis presented here can be extended in various ways. First, it will be interesting to look for the spiky string solution from the D-string in this background and look for the finite-size effect. It is worth investigating the finite-size effect of an oscillating $(m, n)$-string in the mixed flux background. 

\vskip .1in
\noindent
{\bf Acknowledgment:} One of us (RRN) would like to thank the financial support from Department of Science and Technology (DST), Ministry of Science and Technology, Government of India under the Women Scientist Schemen A (WOS-A) (grant no. SR/WOS-A/PM-62/2017).
	\appendix
	\appendixpageoff 
	\section{Useful elliptic integrals and Jacobi elliptic functions }
	This appendix is about the definitions and some basic properties of elliptic integrals and also we write some relevant elliptic integrals which have been used in our paper. The incomplete elliptic integral of first kind is defined as
	\begin{eqnarray}
	\mathbb{F}(\varphi, m)=\int_0^ \varphi \frac{d \theta}{\sqrt{1-m \sin^2 \theta}},
	\end{eqnarray}
	where the range of modulus m and amplitude $\varphi$ are $0\leq m\leq 1$ and $0 \leq \varphi \leq \frac{\pi}{2}$ respectively.
	We write complete elliptic integral when amplitude $\varphi=\frac{\pi}{2}$,
	\begin{equation}
	\mathbb{K}(m)=\mathbb{F}(\frac{\pi}{2}, m).
	\end{equation}
	
	Similarly the elliptic integral of second and third kind are written as
	\begin{eqnarray}
	\mathbb{E}(\varphi,m)&=&\int_{0}^{\varphi} d\theta\sqrt{1-m \sin^2 \theta},~~~~~~~~ \mathbb{E}(m)=\mathbb{E}(\frac{\pi}{2},m), \nonumber\\
	\boldmath{\Pi}(n,\varphi,m)&=&\int_{0}^{\varphi} d\theta\frac{1}{(1-n \sin^2 \theta)\sqrt{1-m \sin^2 \theta}},~~~~~ \boldmath{\Pi}(n,m)=\boldmath{\Pi}(n,\frac{\pi}{2},m). \nonumber\\
	\end{eqnarray}
	Here are some formulas, we need for our calculation
	\begin{eqnarray}
	\int_{z_{min}}^{z_{max}} dz \frac{1}{\sqrt{(z_{max}^2-z^2)(z^2-z_{min}^2)}}=\frac{1}{z_{max}}\mathbb{K}\left(1-\frac{z_{min}^2}{z_{max}^2}\right), \nonumber \\
	\int_{z_{min}}^{z_{max}} dz \frac{z^2}{\sqrt{(z_{max}^2-z^2)(z^2-z_{min}^2)}}=z_{max}\mathbb{E}\left(1-\frac{z_{min}^2}{z_{max}^2}\right), \nonumber \\
	\int_{z_{min}}^{z_{max}} dz \frac{1}{(1-z^2)\sqrt{(z_{max}^2-z^2)(z^2-z_{min}^2)}}=\frac{1}{z_{max}(1-z_{max}^2)}\boldmath{\Pi}\left(\frac{z_{max}^2-z_{min}^2}{z_{max}^2-1},1-\frac{z_{min}^2}{z_{max}^2}\right),\nonumber \\
	\mathbb{F}(\varphi,m)=\frac{1}{\sqrt{m}}\mathbb{F}(\varphi_1,\frac{1}{m})~~~\text{where}~ \varphi_1=\arcsin(\sqrt{m}\sin\varphi).\nonumber\\
	\end{eqnarray}
	\begin{eqnarray}\label{Pi}
	\boldmath{\Pi}(n,m)=\frac{1}{(n-m)\mathbb{K}(1-m)}\left[\frac{\pi}{2}\sqrt{\frac{n(m-n)}{n-1}}\mathbb{F}\left(\arcsin\sqrt{\frac{n-m}{n(1-m)}},1-m\right)- \right. \nonumber \\
	-\left.\mathbb{K}(m)\left[m\mathbb{K}(1-m)-n\boldmath{\Pi}\left(\frac{n-m}{n},1-m\right)\right]\right].\nonumber\\
	\end{eqnarray}
	\section{Lambert $\mathbf{W}$-function}
In this appendix, we briefly discuss Lambert $\mathbf{W}$-function which has been used throughout our paper to present the result. The multi-valued Lambert $\mathbf{W}$-function is defined by the following relation
	\begin{eqnarray}
	\mathbf{W}(z)e^{\mathbf{W}(z)}=z\Leftrightarrow\mathbf{W}(ze^z)=z.
	\end{eqnarray}
	The Lambert $\mathbf{W}$-function has two branches on real line i.e $\mathbf{W}_0(x)$ and $\mathbf{W}_{-}(x)$, the complex variable $z$ replaced by real $x$. For $x\ge 0$, there is only one real branch but for $-1/e \le x <0$, $\mathbf{W}$-function is double-valued. In this interval, 
	the branch satisfying $\mathbf{W}(x)\ge -1$ denoted by $\mathbf{W}_0(x)$ and $\mathbf{W}(x)\le -1$ by $\mathbf{W}_{-1}(x)$. The principal branch $\mathbf{W}_0(x)>0$ for $x>0$ and $\mathbf{W}_0(0)=0$. The lower branch $\mathbf{W}_{-1}(x)$ takes values in $\big(-\infty, -1\big]$ for $x\in \big[-1/e,0\big)$. The following formula gives the asymptotics for the $\mathbf{W}_{-1}\big(x\big)$ both at $x=0$ and at $x=\infty$ \cite{Corless} 
	\begin{eqnarray}
	\mathbf{W}_{-1}\big(x\big)=\ln\abs x -\ln\abs{\ln \abs x}+\sum_{n=0}^{\infty}\sum_{m=1}^{\infty}\frac{(-1)^n}{m!}\left[\begin{array}{cc} n+m \\
	n+1 \end{array}\right] \left(\ln\abs x\right)^{-n-m} \left(\ln\abs{\ln \abs x}\right)^m\nonumber \\
	\end{eqnarray}
	where $\left[\begin{array}{cc} n+m \\
	n+1 \end{array}\right]$, the Stirling number of the first kind can be defined as
	\begin{eqnarray}
		\begin{bmatrix}
	n \\ m \end{bmatrix}=\begin{bmatrix}
	n-1 \\ m-1 \end{bmatrix}+(n-1)	\begin{bmatrix}
	n-1 \\ m \end{bmatrix} \quad \text{and} \quad 	\begin{bmatrix}
	n \\ 0 \end{bmatrix}=	\begin{bmatrix}
	0 \\ m \end{bmatrix}=0, \, 	\begin{bmatrix}
	0 \\ 0 \end{bmatrix}=1, \, n,m\ge 1.\nonumber\\
	\end{eqnarray}

\providecommand{\href}[2]{#2}\begingroup\raggedright\endgroup

\begin{thebibliography}{10}

		\bibitem{Maldacena:1997re}
		J.~M.~Maldacena,
		``The Large N limit of superconformal field theories and supergravity,''
		Adv.\ Theor.\ Math.\ Phys.\  {\bf 2}, 231 (1998)
		[Int.\ J.\ Theor.\ Phys.\  {\bf 38}, 1113 (1999)]
		[hep-th/9711200].
		
		
		\bibitem{Witten:1998qj}
		E.~Witten,
		``Anti-de Sitter space and holography,''
		Adv.\ Theor.\ Math.\ Phys.\  {\bf 2}, 253 (1998)
		[hep-th/9802150].
		
		
		\bibitem{Gubser:1998bc} 
		S.~S.~Gubser, I.~R.~Klebanov and A.~M.~Polyakov,
		``Gauge theory correlators from noncritical string theory,''
		Phys.\ Lett.\ B {\bf 428}, 105 (1998)
		[hep-th/9802109].
		
	\bibitem{Gubser:2002tv}
	S.~S.~Gubser, I.~R.~Klebanov and A.~M.~Polyakov,
	``A Semiclassical limit of the gauge / string correspondence,''
	Nucl.\ Phys.\ B {\bf 636}, 99 (2002)
	[hep-th/0204051].
	
	\bibitem{Kruczenski:2004wg} 
	M.~Kruczenski,
	``Spiky strings and single trace operators in gauge theories,''
	JHEP {\bf 0508}, 014 (2005)
	[hep-th/0410226].
	
	\bibitem{Kruczenski:2006pk} 
	M.~Kruczenski, J.~Russo and A.~A.~Tseytlin,
	``Spiky strings and giant magnons on S**5,''
	JHEP {\bf 0610}, 002 (2006)
	[hep-th/0607044].
	\bibitem{Kruczenski:2008bs} 
	M.~Kruczenski and A.~A.~Tseytlin,
	``Spiky strings, light-like Wilson loops and pp-wave anomaly,''
	Phys.\ Rev.\ D {\bf 77}, 126005 (2008)
	[arXiv:0802.2039 [hep-th]].

	\bibitem{Jevicki:2008mm} 
	A.~Jevicki and K.~Jin,
	``Solitons and AdS String Solutions,''
	Int.\ J.\ Mod.\ Phys.\ A {\bf 23}, 2289 (2008)
	[arXiv:0804.0412 [hep-th]].
	

	
	\bibitem{Ishizeki:2008tx} 
	R.~Ishizeki, M.~Kruczenski, A.~Tirziu and A.~A.~Tseytlin,
	``Spiky strings in AdS(3) x S1 and their AdS-pp-wave limits,''
	Phys.\ Rev.\ D {\bf 79}, 026006 (2009)
	[arXiv:0812.2431 [hep-th]].
	\bibitem{Jevicki:2009uz} 
	A.~Jevicki and K.~Jin,
	``Moduli Dynamics of AdS(3) Strings,''
	JHEP {\bf 0906}, 064 (2009)
	[arXiv:0903.3389 [hep-th]].
	\bibitem{Kruczenski:2010xs} 
	M.~Kruczenski and A.~Tirziu,
	``Spiky strings in Bethe Ansatz at strong coupling,''
	Phys.\ Rev.\ D {\bf 81}, 106004 (2010)
	[arXiv:1002.4843 [hep-th]].
	
	\bibitem{Banerjee:2015nha} 
	A.~Banerjee, S.~Bhattacharya and K.~L.~Panigrahi,
	``Spiky strings in $\varkappa$-deformed $AdS$,''
	JHEP {\bf 1506}, 057 (2015)
	[arXiv:1503.07447 [hep-th]].
	
	\bibitem{Dorey:2008vp} 
	N.~Dorey and M.~Losi,
	``Spiky Strings and Spin Chains,''
	arXiv:0812.1704 [hep-th].
	
	\bibitem{Dorey:2010iy} 
	N.~Dorey and M.~Losi,
	``Giant Holes,''
	J.\ Phys.\ A {\bf 43}, 285402 (2010)
	[arXiv:1001.4750 [hep-th]].
	
	\bibitem{Dorey:2010id} 
	N.~Dorey and M.~Losi,
	``Spiky Strings and Giant Holes,''
	JHEP {\bf 1012}, 014 (2010),
	[arXiv:1008.5096 [hep-th]].
	
	\bibitem{Dorey:2010zz} 
	N.~Dorey and M.~Losi,
	``Spiky strings and gauge theory partons,''
	Int.\ J.\ Mod.\ Phys.\ A {\bf 25}, 4641 (2010).
		
	\bibitem{Hoare:2013pma} 
	B.~Hoare and A.~A.~Tseytlin,
	``On string theory on $AdS_3 x S^3 x T^4$ with mixed 3-form flux: tree-level S-matrix,''
	Nucl.\ Phys.\ B {\bf 873}, 682 (2013),
	[arXiv:1303.1037 [hep-th]].	
	
	\bibitem{Hoare:2013ida} 
	B.~Hoare and A.~A.~Tseytlin,
	``Massive S-matrix of $AdS_3 x S^3 x T^4$ superstring theory with mixed 3-form flux,''
	Nucl.\ Phys.\ B {\bf 873}, 395 (2013),
	[arXiv:1304.4099 [hep-th]].
	
	\bibitem{OhlssonSax:2018hgc} 
	O.~Ohlsson Sax and B.~Stefa\'nski,
	``Closed strings and moduli in AdS$_{3}$/CFT$_{2}$,''
	JHEP {\bf 1805}, 101 (2018),
	[arXiv:1804.02023 [hep-th]].
	
	\bibitem{Lloyd:2014bsa} 
	T.~Lloyd, O.~Ohlsson Sax, A.~Sfondrini and B.~Stefa\'nski, Jr.,
	``The complete worldsheet S matrix of superstrings on $AdS_3 \times S^3 \times T^4$ with mixed three-form flux,''
	Nucl.\ Phys.\ B {\bf 891}, 570 (2015)
	[arXiv:1410.0866 [hep-th]].
	\bibitem{Borsato:2015mma} 
	R.~Borsato, O.~Ohlsson Sax, A.~Sfondrini and B.~Stefa\'nski,
	``The $\mathrm{AdS}_3\times \mathrm{S}^3\times \mathrm{S}^3\times\mathrm{S}^1$ worldsheet S matrix,''
	J.\ Phys.\ A {\bf 48}, no. 41, 415401 (2015),
	[arXiv:1506.00218 [hep-th]].
	
	\bibitem{Baggio:2018gct} 
	M.~Baggio and A.~Sfondrini,
	``Strings on NS-NS Backgrounds as Integrable Deformations,''
	Phys.\ Rev.\ D {\bf 98}, no. 2, 021902 (2018)
	doi:10.1103/PhysRevD.98.021902
	[arXiv:1804.01998 [hep-th]].
	
	\bibitem{Sfondrini:2014via} 
	A.~Sfondrini,
	``Towards integrability for ${\rm Ad}{{{\rm S}}_{{\bf 3}}}/{\rm CF}{{{\rm T}}_{{\bf 2}}}$,''
	J.\ Phys.\ A {\bf 48}, no. 2, 023001 (2015),
	[arXiv:1406.2971 [hep-th]].
	
	\bibitem{Dei:2018jyj} 
	A.~Dei and A.~Sfondrini,
	JHEP {\bf 1902}, 072 (2019)
	doi:10.1007/JHEP02(2019)072
	[arXiv:1812.08195 [hep-th]].
	\bibitem{Dei:2018mfl} 
	A.~Dei and A.~Sfondrini,
	``Integrable spin chain for stringy Wess-Zumino-Witten models,''
	JHEP {\bf 1807}, 109 (2018)
	[arXiv:1806.00422 [hep-th]].
	
	\bibitem{Bombardelli:2018jkj} 
	D.~Bombardelli, B.~Stefa\'nski and A.~Torrielli,
	``The low-energy limit of AdS$_{3}$/CFT$_{2}$ and its TBA,''
	JHEP {\bf 1810}, 177 (2018)
	[arXiv:1807.07775 [hep-th]].
		
	
	\bibitem{Fontanella:2019ury} 
	A.~Fontanella, O.~Ohlsson Sax, B.~Stefa\'nski, Jr. and A.~Torrielli,
	``The effectiveness of relativistic invariance in AdS$_{3}$,''
	JHEP {\bf 1907}, 105 (2019)
	doi:10.1007/JHEP07(2019)105
	[arXiv:1905.00757 [hep-th]].
	
	\bibitem{Pittelli:2014ria} 
	A.~Pittelli, A.~Torrielli and M.~Wolf,
	``Secret symmetries of type IIB superstring theory on $AdS_3 \times S^3 \times M^4$,''
	J.\ Phys.\ A {\bf 47}, no. 45, 455402 (2014)
	doi:10.1088/1751-8113/47/45/455402
	[arXiv:1406.2840 [hep-th]].
	
	\bibitem{Pittelli:2017spf} 
	A.~Pittelli,
	``Yangian Symmetry of String Theory on $AdS_3 \times S^3 \times S^3 \times S^1$ with Mixed 3-form Flux,''
	Nucl.\ Phys.\ B {\bf 935}, 271 (2018)
	doi:10.1016/j.nuclphysb.2018.08.013
	[arXiv:1711.02468 [hep-th]].
	
	\bibitem{Hoare:2013lja} 
	B.~Hoare, A.~Stepanchuk and A.~A.~Tseytlin,
	``Giant magnon solution and dispersion relation in string theory in $AdS_3$x$S^3$x$T^4$ with mixed flux,''
	Nucl.\ Phys.\ B {\bf 879}, 318 (2014),
	[arXiv:1311.1794 [hep-th]].
	
	\bibitem{Lee:2008sk} 
	B.~H.~Lee, R.~R.~Nayak, K.~L.~Panigrahi and C.~Park,
	``On the giant magnon and spike solutions for strings on AdS(3) x S**3,''
	JHEP {\bf 0806}, 065 (2008),
	[arXiv:0804.2923 [hep-th]].
	
	\bibitem{David:2014qta} 
	J.~R.~David and A.~Sadhukhan,
	``Spinning strings and minimal surfaces in $AdS_3$ with mixed 3-form fluxes,''
	JHEP {\bf 1410}, 49 (2014),
	[arXiv:1405.2687 [hep-th]].
	
\bibitem{Banerjee:2014gga} 
A.~Banerjee, K.~L.~Panigrahi and P.~M.~Pradhan,
``Spiky strings on $AdS_3 \times S^3$ with NS-NS flux,''
Phys.\ Rev.\ D {\bf 90}, no. 10, 106006 (2014),
[arXiv:1405.5497 [hep-th]].
	
\bibitem{Banerjee:2015qeq} 
A.~Banerjee and A.~Sadhukhan,
``Multi-spike strings in AdS$_{3}$ with mixed three-form fluxes,''
JHEP {\bf 1605}, 083 (2016),
[arXiv:1512.01816 [hep-th]].	
	
\bibitem{Minahan:2002ve} 
J.~A.~Minahan and K.~Zarembo,
``The Bethe ansatz for N=4 superYang-Mills,''
JHEP {\bf 0303}, 013 (2003),
[hep-th/0212208].

\bibitem{Beisert:2003jj} 
N.~Beisert,
``The complete one loop dilatation operator of N=4 superYang-Mills theory,''
Nucl.\ Phys.\ B {\bf 676}, 3 (2004),
[hep-th/0307015].

\bibitem{Beisert:2003yb} 
N.~Beisert and M.~Staudacher,
``The N=4 SYM integrable super spin chain,''
Nucl.\ Phys.\ B {\bf 670}, 439 (2003),
[hep-th/0307042].

\bibitem{Beisert:2004hm} 
N.~Beisert, V.~Dippel and M.~Staudacher,
``A Novel long range spin chain and planar N=4 super Yang-Mills,''
JHEP {\bf 0407}, 075 (2004),
[hep-th/0405001].
	
\bibitem{Sieg:2005kd} 
C.~Sieg and A.~Torrielli,
``Wrapping interactions and the genus expansion of the 2-point function of composite operators,''
Nucl.\ Phys.\ B {\bf 723}, 3 (2005)
[hep-th/0505071].

\bibitem{Ambjorn:2005wa} 
J.~Ambjorn, R.~A.~Janik and C.~Kristjansen,
``Wrapping interactions and a new source of corrections to the spin-chain/string duality,''
Nucl.\ Phys.\ B {\bf 736}, 288 (2006)
[hep-th/0510171].

\bibitem{Kotikov:2007cy} 
A.~V.~Kotikov, L.~N.~Lipatov, A.~Rej, M.~Staudacher and V.~N.~Velizhanin,
``Dressing and wrapping,''
J.\ Stat.\ Mech.\  {\bf 0710}, P10003 (2007)
[arXiv:0704.3586 [hep-th]].

\bibitem{Gromov:2009tv} 
N.~Gromov, V.~Kazakov and P.~Vieira,
``Exact Spectrum of Anomalous Dimensions of Planar N=4 Supersymmetric Yang-Mills Theory,''
Phys.\ Rev.\ Lett.\  {\bf 103}, 131601 (2009)
[arXiv:0901.3753 [hep-th]].

\bibitem{Gromov:2013pga} 
N.~Gromov, V.~Kazakov, S.~Leurent and D.~Volin,
``Quantum Spectral Curve for Planar $\mathcal{N} = 4$ Super-Yang-Mills Theory,''
Phys.\ Rev.\ Lett.\  {\bf 112}, no. 1, 011602 (2014)
[arXiv:1305.1939 [hep-th]].

\bibitem{SchaferNameki:2005tn} 
S.~Schafer-Nameki, M.~Zamaklar and K.~Zarembo,
``Quantum corrections to spinning strings in AdS(5) x S(5) and Bethe ansatz: A Comparative study,''
JHEP {\bf 0509}, 051 (2005)
[hep-th/0507189].
	
\bibitem{SchaferNameki:2006gk} 
S.~Schafer-Nameki,
``Exact expressions for quantum corrections to spinning strings,''
Phys.\ Lett.\ B {\bf 639}, 571 (2006)
[hep-th/0602214].


\bibitem{SchaferNameki:2006ey} 
S.~Schafer-Nameki, M.~Zamaklar and K.~Zarembo,
``How Accurate is the Quantum String Bethe Ansatz?,''
JHEP {\bf 0612}, 020 (2006)
[hep-th/0610250].


\bibitem{Arutyunov:2006gs} 
G.~Arutyunov, S.~Frolov and M.~Zamaklar,
``Finite-size Effects from Giant Magnons,''
Nucl.\ Phys.\ B {\bf 778}, 1 (2007)
[hep-th/0606126].


\bibitem{Astolfi:2007uz} 
D.~Astolfi, V.~Forini, G.~Grignani and G.~W.~Semenoff,
``Gauge invariant finite size spectrum of the giant magnon,''
Phys.\ Lett.\ B {\bf 651}, 329 (2007)
[hep-th/0702043 [HEP-TH]].

\bibitem{Klose:2008rx} 
T.~Klose and T.~McLoughlin,
``Interacting finite-size magnons,''
J.\ Phys.\ A {\bf 41}, 285401 (2008)
[arXiv:0803.2324 [hep-th]].


\bibitem{Ramadanovic:2008qd} 
B.~Ramadanovic and G.~W.~Semenoff,
``Finite Size Giant Magnon,''
Phys.\ Rev.\ D {\bf 79}, 126006 (2009)
[arXiv:0803.4028 [hep-th]].


\bibitem{Minahan:2008re} 
J.~A.~Minahan and O.~Ohlsson Sax,
``Finite size effects for giant magnons on physical strings,''
Nucl.\ Phys.\ B {\bf 801}, 97 (2008)
[arXiv:0801.2064 [hep-th]].

\bibitem{Shenderovich:2008bs} 
I.~Shenderovich,
``Giant magnons in AdS(4) / CFT(3): Dispersion, quantization and finite-size corrections,''
arXiv:0807.2861 [hep-th].

\bibitem{Ahn:2008gd} 
C.~Ahn and P.~Bozhilov,
``Finite-size effects of Membranes on AdS(4) x S**7,''
JHEP {\bf 0808}, 054 (2008)
[arXiv:0807.0566 [hep-th]].


\bibitem{Jain:2008mt} 
S.~Jain and K.~L.~Panigrahi,
``Spiky Strings in AdS(4) x CP**3 with Neveu-Schwarz Flux,''
JHEP {\bf 0812}, 064 (2008)
[arXiv:0810.3516 [hep-th]].

\bibitem{Hatsuda:2008na} 
Y.~Hatsuda and R.~Suzuki,
JHEP {\bf 0809}, 025 (2008)
[arXiv:0807.0643 [hep-th]].

\bibitem{Ahn:2008sk} 
C.~Ahn and P.~Bozhilov,
``Finite-size Effects for Single Spike,''
JHEP {\bf 0807}, 105 (2008)
[arXiv:0806.1085 [hep-th]].


\bibitem{Grignani:2008te} 
G.~Grignani, T.~Harmark, M.~Orselli and G.~W.~Semenoff,
``Finite size Giant Magnons in the string dual of N=6 superconformal Chern-Simons theory,''
JHEP {\bf 0812}, 008 (2008)
[arXiv:0807.0205 [hep-th]].

\bibitem{Bozhilov:2010rf} 
P.~Bozhilov,
``Close to the Giant Magnons,''
arXiv:1010.5465 [hep-th].

\bibitem{Floratos:2014gqa} 
E.~Floratos and G.~Linardopoulos,
``Large-Spin and Large-Winding Expansions of Giant Magnons and Single Spikes,''
Nucl.\ Phys.\ B {\bf 897}, 229 (2015)
[arXiv:1406.0796 [hep-th]].

\bibitem{Gromov:2008ie} 
N.~Gromov, S.~Schafer-Nameki and P.~Vieira,
``Quantum Wrapped Giant Magnon,''
Phys.\ Rev.\ D {\bf 78}, 026006 (2008)
[arXiv:0801.3671 [hep-th]].	

\bibitem{Heller:2008at} 
M.~P.~Heller, R.~A.~Janik and T.~Lukowski,
``A New derivation of Luscher F-term and fluctuations around the giant magnon,''
JHEP {\bf 0806}, 036 (2008)
[arXiv:0801.4463 [hep-th]].

\bibitem{Bykov:2008bj} 
D.~V.~Bykov and S.~Frolov,
``Giant magnons in TsT-transformed AdS(5) x S**5,''
JHEP {\bf 0807}, 071 (2008)
[arXiv:0805.1070 [hep-th]].


\bibitem{Ahn:2010da} 
C.~Ahn and P.~Bozhilov,
``Finite-Size Dyonic Giant Magnons in TsT-transformed $AdS_5\times S^5$,''
JHEP {\bf 1007}, 048 (2010)
[arXiv:1005.2508 [hep-th]].


\bibitem{Ahn:2014aqa} 
C.~Ahn and P.~Bozhilov,
``Finite-size giant magnons on $\eta$-deformed $AdS_5 \times S^5$,''
Phys.\ Lett.\ B {\bf 737}, 293 (2014)
[arXiv:1406.0628 [hep-th]].


\bibitem{Ahn:2016egk} 
C.~Ahn,
``Finite-size effect of $\eta$-deformed $AdS_5 \times S^5$ at strong coupling,''
Phys.\ Lett.\ B {\bf 767}, 121 (2017)
[arXiv:1611.09992 [hep-th]].


\bibitem{Panigrahi:2018xuv} 
K.~L.~Panigrahi and M.~Samal,
``Finite Size Effect from Classical Strings in deformed AdS$_3\times$ S$^3$,''
JHEP {\bf 1809}, 162 (2018)
[arXiv:1807.04601 [hep-th]].
	
\bibitem{Barik:2019uqf} 
S.~P.~Barik and K.~L.~Panigrahi,
``Finite size effect from classical strings in $AdS_3 \times S^3$ with NS-NS flux,''
arXiv:1901.03026 [hep-th].

\bibitem{Floratos:2013cia} 
E.~Floratos, G.~Georgiou and G.~Linardopoulos,
``Large-Spin Expansions of GKP Strings,''
JHEP {\bf 1403}, 018 (2014)
[arXiv:1311.5800 [hep-th]].	
	
\bibitem{Corless}
R.M. Corless, G.H. Gonnet, D.E.G. Hare, D.J. Jeffrey, and D.E. Knuth,
``On the Lambert W Function,'' Adv Comput Math (1996) 5: 329. 

		
\end{thebibliography}
\end{document}